%% file: menendez_nuphys15.tex
\documentclass[12pt]{article}
\usepackage{graphicx}
\usepackage{amsmath,amssymb}
\usepackage[english]{babel}
\usepackage[latin1]{inputenc}
\usepackage{times}
\usepackage[T1]{fontenc}
\usepackage{amsmath,amsfonts,amssymb,bm}
\usepackage{bbold,nicefrac}

\def\todai{Department of Physics\\
The University of Tokyo, Bunkyo-ku 113-0033, Tokyo, Japan}
\def\support{\footnote{Work supported by 
an International Research Fellowship of the Japan Society for the Promotion of Science (JSPS),
and Grant-in-Aid for Scientific Research No. 26$\cdot$04323.
}}

\def\0nbb{$0\nu\beta\beta$}

\def\Title#1{\begin{center} {\Large #1 } \end{center}}
\def\Author#1{\begin{center}{ \sc #1} \end{center}}
\def\Address#1{\begin{center}{ \it #1} \end{center}}

\newenvironment{Abstract}{\begin{quotation}  }{\end{quotation}}
\newenvironment{Presented}{\begin{quotation} \begin{center} 
             PRESENTED AT\end{center}\bigskip 
      \begin{center}\begin{large}}{\end{large}\end{center} \end{quotation}}
\def\Acknowledgements{\bigskip  \bigskip \begin{center} \begin{large}
             \bf ACKNOWLEDGEMENTS \end{large}\end{center}}

\input econfmacros.tex

\begin{document}
\begin{titlepage}

\vfill
\Title{What do we know about neutrinoless double-beta decay nuclear matrix elements?}
\vfill
\Author{ Javier Men\'{e}ndez\support}
\Address{\todai}
\vfill
\begin{Abstract}
The detection of neutrinoless double-beta decay will establish the Majorana nature of neutrinos.
In addition, if the nuclear matrix elements of this process are reliably known,
the experimental lifetime will provide precious information about the absolute neutrino masses and 
hierarchy. I review the status of nuclear structure calculations for neutrinoless double-beta decay
matrix elements, and discuss some key issues to be addressed in order to meet the demand for accurate nuclear matrix elements.
\end{Abstract}
\vfill
\begin{Presented}
NuPhys2015, Prospects in Neutrino Physics \\
Barbican Centre, London, UK,  December 16--18, 2015
\end{Presented}
\vfill
\end{titlepage}
\def\thefootnote{\fnsymbol{footnote}}
\setcounter{footnote}{0}

\section{Neutrinoless double-beta decay}
\label{0nbb}

Neutrinoless double-beta (\0nbb) decay is a very special process.
Most importantly, the experimental detection of this lepton-number violating decay
will proof the Majorana nature of neutrinos,
this is, that they are their own antiparticle. 
In addition, the lifetime of the \0nbb decay
is related to the neutrino masses so that its measurement
will also probe the unknown absolute neutrino mass and hierarchy.

However, there is yet another ingredient in the connection
between the \0nbb decay lifetime and the neutrino mass:
since it is a nuclear decay, the lifetime naturally depends on
the nuclear matrix element (NME) of the transition.
Overall, the \0nbb decay half-life can be written as~\cite{08avignone}
\begin{equation}  
\left[T_{1/2}^{0 \nu\beta\beta}\left(0^+_i\rightarrow 0^+_f\right)\right]^{-1}
= G^{0 \nu\beta\beta} |M^{0 \nu\beta\beta}|^2  m_{\beta \beta}^2,  
\end{equation}
with $M^{0 \nu\beta\beta}$ the NME, $m_{\beta \beta}$
a combination of the absolute neutrino masses and the neutrino mixing matrix,
and $G^{0\nu\beta\beta}$ a well-known phase-space factor~\cite{12kotila}.
It is apparent, therefore, that for \0nbb decay experiments
~\cite{14exo,13kamlandzen,13gerda,15cuore} to be able to unveil the neutrino masses,
the NMEs of the decay have to be accurately known. Is this presently the case?

To answer this question, let us recall that the NME
in the closure approximation is~\cite{08avignone}
\begin{equation}
 M^{0\nu\beta\beta} = \left\langle 0_{f}^{+}\right|
\hat{O}^{0\nu\beta\beta}
 \left|0_{i}^{+}\right\rangle.
\end{equation}
Therefore, a reliable NME relies on two independent parts:
first, the nuclear structure of the transition initial and final states;
second, the evaluation of the decay operator $\hat{O}^{0\nu\beta\beta}$ between these states.
In the following, these two parts are analysed separately.

\section{The initial and final nuclei: nuclear structure}
\label{sec:structure}

\begin{figure}[t]
\centering
\includegraphics[width=\textwidth]{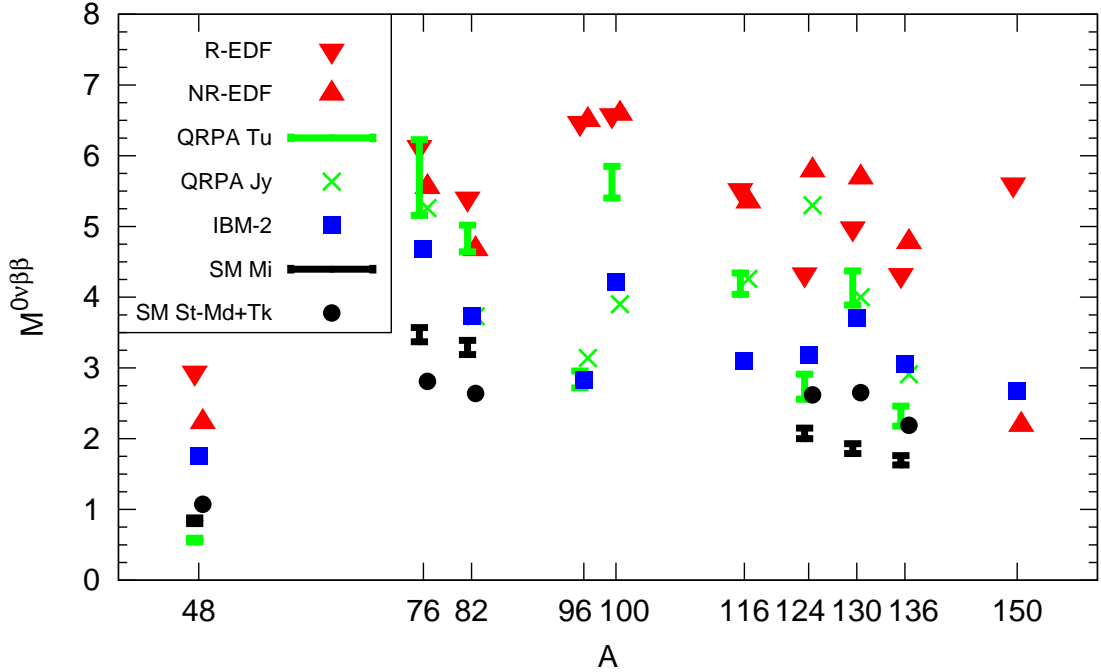}
\caption{
NMEs for the \0nbb decays of $^{48}$Ca, $^{76}$Ge, $^{82}$Se, $^{96}$Zr, $^{100}$Mo,
$^{116}$Cd, $^{124}$Sn, $^{130}$Te, $^{136}$Xe and $^{150}$Nd,
shown according to their mass number $A$.
Results are shown for the shell model calculations of the
Strasbourg-Madrid~\cite{09menendez} and Tokyo~\cite{16iwata} groups (SM St-Ma+Tk),
and the Michigan group~\cite{16neacsu} (SM Mi),
the interacting boson model~\cite{15barea} (IBM-2),
the quasiparticle random-phase approximation approach of the Jyv\"askyl\"a~\cite{15hyvarinen}
(QRPA Jy) and T\"ubingen~\cite{13simkovic} (QRPA Tu) groups,
and the non-relativistic~\cite{13vaquero} and relativistic~\cite{15yao}
energy density functional frameworks (NR-EDF and R-EDF, respectively).
}
\label{fig:compare_nmes}
\end{figure}

First let us focus on
the impact of the nuclear structure of the initial and final states in the \0nbb decay NMEs.
Very different nuclear structure approaches have been used to study this process. 
Figure~\ref{fig:compare_nmes} shows an updated comparison of the main NME calculations
obtained with various nuclear structure frameworks
~\cite{09menendez,16iwata,16neacsu,15barea,13simkovic,15hyvarinen,13vaquero,15yao}.
The differences are about a factor of two to three, or three to four units.

Among the smallest NMEs are those from shell model calculations.
The nuclear shell model is very successful in describing
nuclear masses, low-lying excited states, electromagnetic transitions and single-$\beta$ decays
over a wide range of nuclei~\cite{05caurier}.
These calculations can include very rich nuclear structure correlations.
However, they are typically performed in a rather limited configuration space
of one major harmonic-oscillator shell,
while the remaining orbitals are taken into account only perturbatively.

\begin{figure}[t]
\centering
\includegraphics[angle=-90,width=\textwidth]{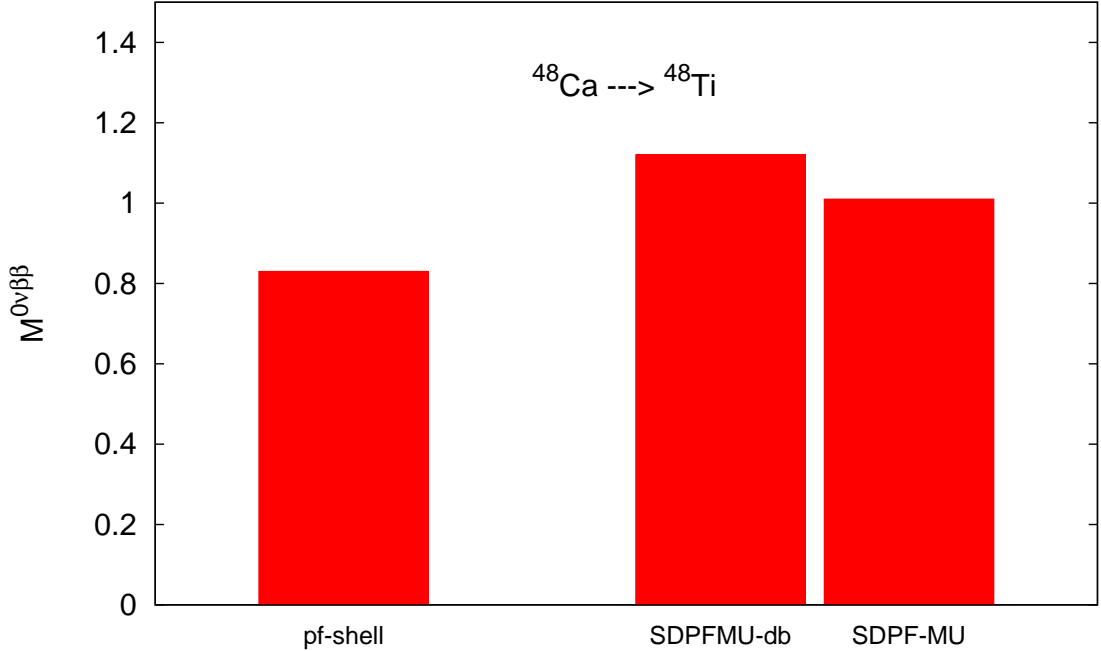}
\caption{
Shell model NME for the \0nbb decay of $^{48}$Ca in one major harmonic-oscillator shell (left),
and in two calculations in two major harmonic-oscillator shells (right), from Ref.~\cite{16iwata}.
The enhancement of the NME in the enlarged configuration space is about 30\%.
}
\label{fig:2hw}
\end{figure}

In order to explore the impact of the size of the configuration space in shell model NMEs~\cite{12vogel},
a very recent work by the Tokyo group focused on the \0nbb decay of $^{48}$Ca~\cite{16iwata}.
Previous studies used a configuration space comprising the $pf$-shell,
this is, assuming a $^{40}$Ca core with eight neutrons in the four $pf$-shell orbitals
~\cite{09menendez}.
In Ref.~\cite{16iwata}, the configuration space was expanded
to include two major harmonic-oscillator shells, adding the $sd$-shell to the $pf$-shell.
In this case, a core of $^{16}$O was assumed,
allowing up to total $2\hbar\omega$ proton and neutron
cross-shell excitations from the $sd$- into the $pf$-shell.
As a result, the size of the diagonalization needed to describe the daughter nucleus $^{48}$Ti
increases from less than $10^6$ to over $10^9$, at the limit of present capabilities.
The effect of the extended calculation compared the one-major-shell one 
is illustrated in Fig.~\ref{fig:2hw}. The NME increases by about 30\%,
with the enhancement produced by additional cross-shell pairing correlations
incorporated in the enlarged configuration space~\cite{16iwata}.
However, the improved NME is still far from the results of other approaches,
suggesting that the size of the shell model configuration space may not explain
the disagreement between NME calculations.

Another important aspect for \0nbb decay NMEs are nuclear structure correlations.
Actually in some cases the disagreement between NME calculations is strongly reduced when
they are restricted to uncorrelated (and therefore too simplistic) initial and final 
states~\cite{14menendez}.
As already pointed out in the case of the $^{48}$Ca decay,
pairing correlations are very important in this process.
Proton-proton and neutron-neutron pairing correlations favour \0nbb decay:
the more of these correlations in the initial and final states, the larger the NMEs
~\cite{08caurier,13vaquero}.
This explains why the additional pairing correlations captured in the two major-shell calculation
enhance the $^{48}$Ca \0nbb decay NME.
Similarly, if pairing correlations are overestimated, the NMEs will be overpredicted
~\cite{08caurier}.

Proton-neutron pairing correlations (more precisely isoscalar pairing correlations)
also impact \0nbb decay~\cite{86vogel}.
In contrast to like-particle pairing,
neglecting isoscalar pairing results in overpredicted NMEs.
This may be somewhat surprising because proton-neutron pairing correlations are usually not
very relevant in nuclear structure.
However, in single-$\beta$ decays and $\beta\beta$ decays, isoscalar pairing is crucial
because neglecting this term breaks the spin-isospin SU(4) symmetry
of the operators, which are therefore especially sensitive to these correlations.
A proper treatment of isoscalar pairing is important for \0nbb decay
because energy density functional methods
(that predict the largest NMEs as shown in Fig.~\ref{fig:compare_nmes})
and the interacting boson model do not include these correlations explicitly.
Without a dedicated calculation it is difficult to quantify the impact
of isoscalar pairing correlations in the NMEs, but a recent shell model study suggests that
the effect could be as large as a 50\% NME reduction ~\cite{16menendez}.

In addition, quadrupole correlations related to deformation are also relevant
for \0nbb decay~\cite{10rodriguez,11bmenendez}.
In this case, quadrupole correlations reduce the NMEs,
especially when the deformation of the initial and final states is different.
The treatment of deformation may explain the different NMEs
between the two energy density functional calculations for $^{150}$Nd,
the only strongly deformed \0nbb decay candidate.

All NMEs available so far are based on phenomenological nuclear structure calculations.
One of the main advances in low-energy nuclear physics in the recent decade is the capability
of performing first principles calculations
based on the underlying theory of the strong interaction, QCD,
combined with improved many-body methods that make use of state-of-the-art computational resources.
For instance, nuclear structure calculations using interactions derived from
chiral effective field theory (EFT)~\cite{09epelbaum},
an effective theory based on the symmetries of QCD, have been very successful
in describing and predicting properties of medium-mass nuclei up to calcium~\cite{15hebeler}.
Moreover, in selected cases the many-body problem can be solved
with all nucleons explicitly included.
These ab initio approaches are not able to provide \0nbb decay NMEs yet,
but they will be able to do so in the near future.
As a first step, single-$\beta$ decays of medium-mass nuclei,
albeit for isotopes lighter than those relevant for \0nbb decay experiments,
are already available~\cite{14ekstrom}.

\section{The transition operator: two-body corrections}

The different NMEs discussed in Sec.~\ref{sec:structure} assume a common transition operator 
entirely consisting of axial and vector weak one-body (1b) currents.
However, studies of light nuclei with mass number $A \lesssim 10$ manifest the need
to go beyond the 1b level to describe magnetic moments and transitions~\cite{14bacca},
or single-$\beta$ decays~\cite{09gazit}.

\begin{figure}[t]
\centering
\includegraphics[width=\textwidth]{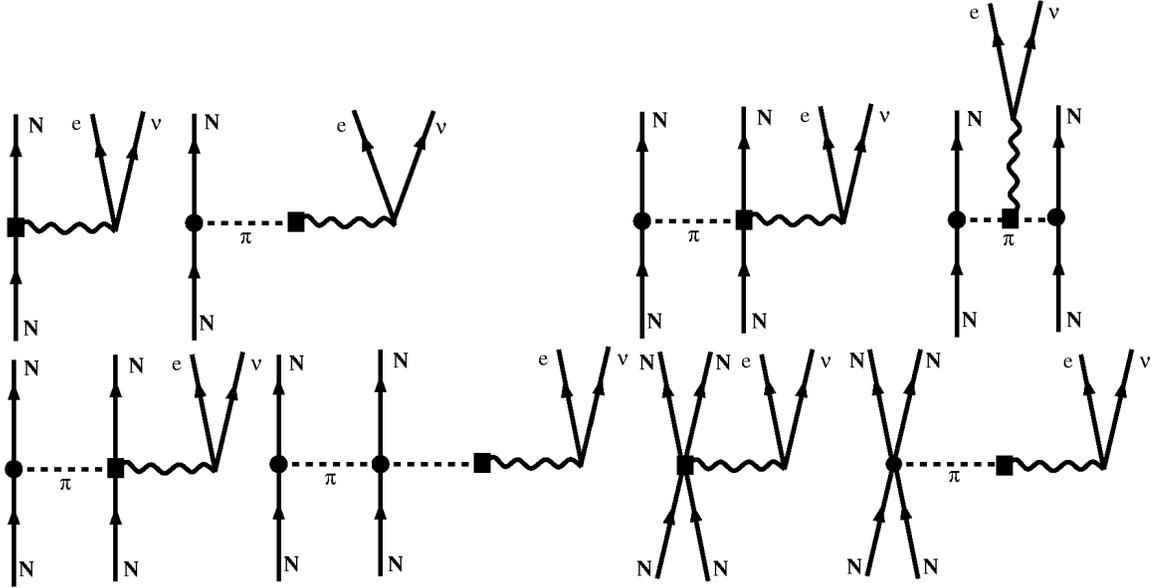}
\caption{
Diagrams corresponding to the 1b currents (upper left part),
vector 2b currents (upper right part) and axial 2b currents (lower part)
relevant for \0nbb decay. 
}
\label{fig:diagrams}
\end{figure}

Chiral EFT, in addition to a theory of nuclear interactions, also predicts how nucleons
interact with external probes, in particular via the weak interaction.
Since chiral EFT is an effective theory,
different terms are organized in orders in the expansion coefficient $Q$.
Chiral EFT predicts that two-body (2b), or meson-exchange currents
enter \0nbb decay at order $Q^2$ in the vector current 
and at order $Q^3$ in the axial current~\cite{15hoferichter}.
This is important because the 1b terms used in standard \0nbb decay calculations
correspond to 1b currents to order $Q^2$,
and the next 1b current contributions only appear at order $Q^4$.
Figure~\ref{fig:diagrams} schematically shows the diagrams of the leading 1b and 2b currents
to order $Q^3$.

The 1b terms relevant for \0nbb decay to order $Q^2$ are~\cite{11menendez}
\begin{align}
V_{\text{1b}}=\tau^-g_V(q), \qquad
\bm{V}_{\text{1b}}&=\tau^-
\left[(1+g_M) \frac{\left(-i\bm{\sigma} \times \bm{q}\right)}{2M} \right],  \nonumber \\
\bm{A}_{\text{1b}}&=\tau^- 
\left[g_A (q) \bm{\sigma} - g_P (q)
\left(\bm{q}\cdot\bm{\sigma} \right) \bm{q}  \right],
\label{1bc}
\end{align}
with $M$ the nucleon mass and ${\bm q}$ the momentum-transfer of the transition.
At vanishing momentum transfer $g_V(0)=1$ because of the conserved vector current,
and $g_A(0)=g_A$, the axial coupling constant.
The coefficient $g_M$ accounts for the isovector anomalous magnetic moment of the nucleon,
and $g_P(q)$ is fixed by the Goldberger-Treiman relation~\cite{11menendez}.

The leading correction to the 1b terms in Eq.~\ref{1bc} are 2b currents.
The evaluation of these 2b terms in \0nbb decay is challenging,
because in general they will lead to a four-body operator.
As an attempt to estimate the importance of 2b effects in \0nbb decay,
the easiest approach is to perform a normal-ordering approximation over a spin-isospin symmetric
reference state (Fermi gas)~\cite{11menendez},
which results in an effective 1b current coming from the 2b terms.
The result can be easily compared to the leading 1b contributions:
\begin{align}
\bm{V}^{\text{NO}}_{\text{2b}}&=\tau^-
\left[\delta m (q) \frac{\left(-i\bm{\sigma} \times \bm{q}\right)}{2M} \right], \nonumber \\
\bm{A}^{\text{NO}}_{\text{2b}}&= \tau^- 
\left[\delta a (q) \bm{\sigma} - \delta p (q)
\left(\bm{q}\cdot\bm{\sigma} \right) \bm{q}  \right], 
\label{2bc_no}
\end{align}
where $\delta a(q)$, $\delta p(q)$ and $\delta m(q)$ can be evaluated
with the low-energy chiral EFT couplings.
Thus, in this approximation the 2b currents amount to a momentum-transfer dependent 
modification of the magnetic, Gamow-Teller and pseudoscalar 1b terms.

\begin{figure}[t]
\centering
\includegraphics[width=\textwidth]{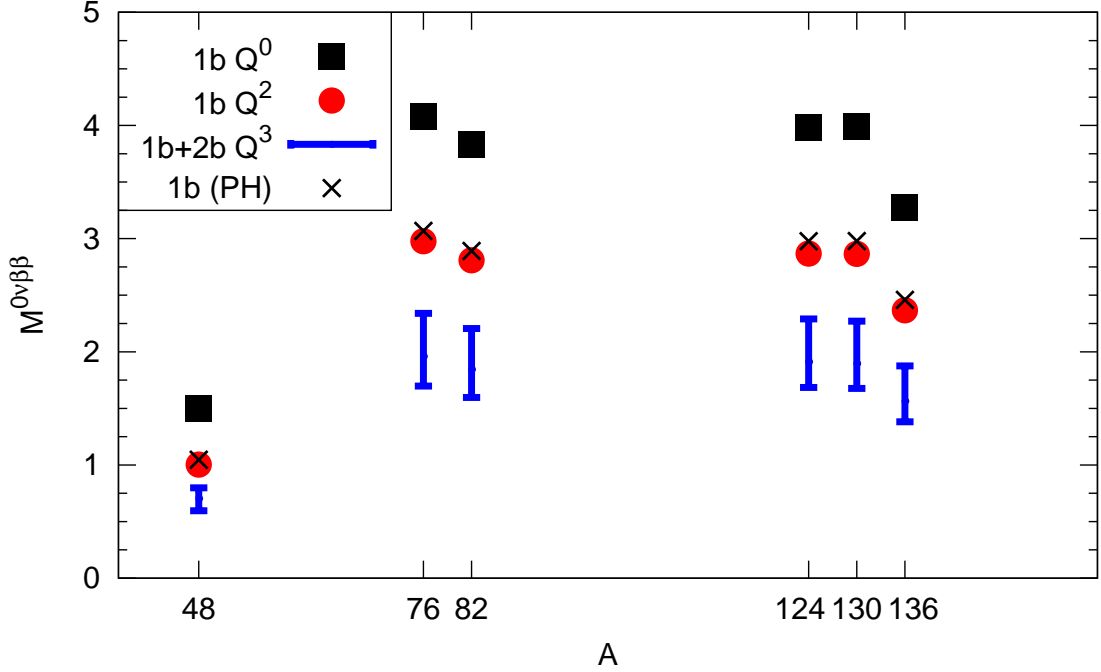}
\caption{
Shell model NMEs for the \0nbb decays of $^{48}$Ca, $^{76}$Ge, $^{82}$Se, $^{124}$Sn, $^{130}$Te and $^{136}$Xe, shown according to their mass number $A$.
The different calculations include chiral 1b currents to leading order $Q^0$ (black squares)
and order $Q^2$ (red circles), and also 2b axial and vector currents (blue bars)
which are the only additional contributions to order $Q^3$.
The NMEs are compared with those obtained with phenomenological 1b currents (black crosses),
roughly corresponding to a $Q^2$ 1b current calculation.
}
\label{fig:2bc_nmes}
\end{figure}

The most important 2b term is the correction to the Gamow-Teller $\tau^-{\bm \sigma}$ term. 
This 2b contribution is enhanced due to the low-lying $\Delta$-isobar excitation~\cite{11menendez},
and its effect is to reduce the strength of the 1b Gamow-Teller term.
A similar need to reduce the strength of this operator is well-known in nuclear structure 
calculations trying to reproduce the experimental lifetime of Gamow-Teller transitions,
a phenomenon usually referred to as Gamow-Teller quenching~\cite{05caurier}.
Even though there may be additional mechanisms leading to this quenching,
such as corrections due to the limitations in the nuclear structure calculations,
the estimation in Ref.~\cite{11menendez}
suggests that 2b currents are a significant contribution.

The other two terms in Eq.~\eqref{2bc_no} are relevant at high momentum transfers.
They also impact \0nbb decay as in this process typically $q\sim200$ MeV
due to the virtual nature of the neutrinos~\cite{12vogel}.
The combined effect of these terms is to partially compensate for the reduction produced
by the leading 2b contribution~\cite{11menendez}.

Figure~\ref{fig:2bc_nmes} shows NMEs
calculated with chiral EFT 1b and 2b currents in the shell model framework.
Only the long-range contributions are included,
because they are expected to be dominant (due to its relation to the $\Delta$-isobar),
and also because the associated chiral EFT couplings
are less precisely known for the short-range parts.
The NMEs are reduced by the 2b corrections in about 35\%,
with a relatively large error band stemming from the uncertainties in the chiral EFT couplings.
Even though the results in Fig.~\ref{fig:2bc_nmes} rely on a simple normal-ordering approximation
they highlight that accurate NME calculations should carefully include 2b currents.

\section{Conclusions}
The \0nbb decay, besides establishing the Majorana nature of neutrinos,
has the potential shed light on the absolute neutrino mass and hierarchy.
For that purpose, it is critical that the associated NMEs are reliably known. 
The most recent calculations show NME differences of about a factor of two or three,
but the main limitations of the calculations,
such as enlarging the shell model configuration space, or including isoscalar pairing correlations 
have been identified and work is in progress to obtain improved NMEs.
In addition, ab initio studies based on chiral EFT will soon became available. 
On the other hand, 2b current corrections to the transition operator are usually neglected,
but their effect could be sizeable and they should be included in NME calculations.

\Acknowledgements
I would like to thank my collaborators
T. Abe, E. Caurier, J. Engel, D. Gazit, N. Hinohara, M. Honma, Y. Iwata, P. Klos,
G. Mart\'{i}nez-Pinedo, F. Nowacki,  T. Otsuka,  A. Poves, T. R. Rodr\'{i}guez, A. Schwenk,
N. Shimizu and Y. Utsuno
for very enlightening discussions and for using our common results for these proceedings.

\end{document}

%% file: econfmacros.tex
\def\beq{\begin{equation}}
\def\eeq#1{\label{#1}\end{equation}}
\def\eeqn{\end{equation}}

\def\beqa{\begin{eqnarray}}
\def\eeqa#1{\label{#1}\end{eqnarray}}
\def\eeqan{\end{eqnarray}}

\let\bar=\overbar

\def\Dslash{\not{\hbox{\kern-4pt $D$}}}
\def\dslash{\not{\hbox{\kern-2pt $\del$}}}

\def\msb{{\bar{\ssstyle M \kern -1pt S}}}




%% file: menendez_nuphys15.bbl
\begin{thebibliography}{99}

\bibitem{08avignone}
F. T. Avignone III, S. R. Elliott and J. Engel, Rev. Mod. Phys. {\bf 80}, 481 (2008).

\bibitem{12kotila}
J. Kotila and F. Iachello, Phys. Rev. C {\bf 85}, 034316 (2012).

\bibitem{14exo}
J. B. Albert~{\it et al.} (EXO Collaboration), Nature {\bf 510}, 229 (2014).

\bibitem{13kamlandzen}
A. Gando~{\it et al.} (KamLAND-Zen Collaboration), Phys. Rev. Lett. {\bf 110}, 062502 (2013).

\bibitem{13gerda}
M. Agostini~{\it et al.} (GERDA Collaboration), Phys. Rev. Lett. {\bf 111}, 122503 (2013).

\bibitem{15cuore}
K. Alfonso~{\it et al.} (CUORE Collaboration), Phys. Rev. Lett. {\bf 115}, 022502 (2015).


\bibitem{09menendez}
J. Men\'{e}ndez, A. Poves, E. Caurier and F. Nowacki, Nucl. Phys. A {\bf 818}, 139 (2009).

\bibitem{16iwata}
Y. Iwata, N. Shimizu, T. Otsuka, Y. Utsuno, J. Men\'{e}ndez, M. Honma and T. Abe,
Phys. Rev. Lett. {\bf 116}, 112502 (2016).

\bibitem{16neacsu}
A. Neacsu and M. Horoi, Phys. Rev. C {\bf 93}, 024308 (2016).

\bibitem{15barea}
J. Barea, J.  Kotila and F. Iachello, Phys. Rev. C {\bf 91}, 034304 (2015).

\bibitem{15hyvarinen}
J. Hyv\"{a}rinen and J. Suhonen, Phys. Rev. C {\bf 87}, 024613 (2015).

\bibitem{13simkovic}
F. \v{S}imkovic, V. Rodin, A. Faessler and P. Vogel, Phys. Rev. C {\bf 87}, 045501 (2013).

\bibitem{13vaquero}
N. L\'{o}pez Vaquero, T. R. Rodr\'{i}guez and J. L. Egido, Phys. Rev. Lett. {\bf 111}, 142501 (2013).

\bibitem{15yao}
J. Yao, L. Song, K. Hagino, P. Ring and J. Meng, Phys. Rev. C {\bf 91}, 024316 (2015).

\bibitem{05caurier}
E. Caurier, G. Mart\'{i}nez-Pinedo, F. Nowacki, A. Poves and A. P. Zuker, Rev. Mod. Phys. {\bf 77}, 427 (2005).

\bibitem{12vogel}
P. Vogel, J. Phys. G: Nucl. Part. Phys. {\bf 39}, 124002 (2012).

\bibitem{14menendez}
J. Men\'endez, T. R. Rodr\'{i}guez, G. Mart\'{i}nez-Pinedo and A. Poves,
Phys. Rev. C {\bf 90}, 024311 (2014).

\bibitem{08caurier}
E. Caurier, F. Nowacki, J. Men\'endez and A. Poves,  Phys. Rev. Lett. {\bf 100}, 052503 (2008).


\bibitem{86vogel}
P. Vogel and M. R. Zirnbauer, Phys. Rev. Lett. {\bf 57}, 3148 (1986).

\bibitem{16menendez}
J. Men\'{e}ndez, N. Hinohara, J. Engel, G. Mart\'{i}nez-Pinedo and T. R. Rodr\'{i}guez,
Phys. Rev. C {\bf 93}, 014305 (2016).

\bibitem{10rodriguez}
T. R. Rodr\'{i}guez and G. Mart\'{i}nez-Pinedo, Phys. Rev. Lett. {\bf 105}, 252503 (2010).

\bibitem{11bmenendez}
J. Men\'{e}ndez, A. Poves, E. Caurier and F. Nowacki, J. Phys. Conf. Ser. {\bf 267}, 012058 (2011).


\bibitem{09epelbaum}
E. Epelbaum, H. W. Hammer, and U.-G. Mei{\ss}ner, Rev. Mod. Phys. {\bf 81}, 1773 (2009).

\bibitem{15hebeler}
K. Hebeler, J. D. Holt, J. Men\'{e}ndez and A. Schwenk,
Annu. Rev. Nucl. Part. Sci. {\bf 65} 457 (2015).

\bibitem{14ekstrom}
A. Ekstr\"{o}m, G. R. Jansen, K. A. Wendt, G. Hagen, T. Papenbrock, S. Bacca, B.~Carlsson and D. Gazit,
Phys. Rev. Lett. {\bf 113}, 262504 (2014).

\bibitem{14bacca}
S. Bacca and S. Pastore, J. Phys. G: Nucl. Part. Phys. {\bf 41}, 123002 (2014).

\bibitem{09gazit}
D. Gazit, S. Quaglioni and P. Navr\'{a}til, Phys. Rev. Lett. {\bf 103}, 102502 (2009).

\bibitem{15hoferichter}
M. Hoferichter, P. Klos and A. Schwenk, Phys. Lett. B {\bf 746}, 410 (2015).

\bibitem{11menendez}
J. Men\'{e}ndez, D. Gazit and A. Schwenk, Phys. Rev. Lett. {\bf 107}, 062501 (2011).




\end{thebibliography}
